# Web Vulnerability Scanners: A Case Study

Angel Rajan, Emre Erturk

Eastern Institute of Technology, Hawke's Bay

**Abstract**

Cloud security is one of the biggest concerns for many companies. The growth in the number and size of websites increases the need for better securing those websites. Manual testing and detection of web vulnerabilities can be very time consuming. Automated Web Vulnerability Scanners (WVS) help with the detection of vulnerabilities in web applications. Acunetix is one of the widely used vulnerability scanners. Acunetix is also easy to implement and to use. The scan results not only provide the details of the vulnerabilities, but also give information about fixing the vulnerabilities. AcuSensor and AcuMonitor (technologies used by Acunetix) help generate more accurate potential vulnerability results. One of the purposes of this paper is to orient current students of computer security with using vulnerability scanners. Secondly, this paper provides a literature review related to the topic of security vulnerability scanners. Finally, web vulnerabilities are addressed from the mobile device and browser perspectives.

**Introduction**

With the advancements in cloud computing, web services, and browser based applications, most business rely on conducting their business communications and transactions online. However, these websites and web applications are not completely secure. Around 30,000 websites are being attacked every day (Lyne, 2013), and one out of every three websites is vulnerable to hacking (Schupak, 2015). Moreover, ninety percent of passwords are vulnerable to being stolen (Warman, 2013). The increasing number of websites and online applications increases the urgency of securing these websites.

Web security scanners are automated tools that check out websites or web applications for security vulnerabilities, without accessing the application's source code (Saeed, 2014). Web vulnerability scanners help to find vulnerabilities of web applications and websites. A security vulnerability is a weakness that may be exploited to cause damage, but its presence does not cause harm by itself (Jeeva, Raveena, Sangeetha, & Vinothini, 2016). Grabber, Vega, Acunetix, Wapiti (InfoSec Institute, 2014) are few examples of web vulnerability scanners.

The Cloud Security Alliance (2016) has recently identified twelve major types of security concerns and threats. Many of these are relevant to areas where web vulnerability scanners may be helpful in reducing risks. For example, two of these concerns, insecure APIs and insufficient due diligence, may be overlooked by web developers and web masters.

This report focuses on Acunetix, one of the widely used web vulnerability scanner. It begins with a description of Acunetix with the details of Acunetix used for this case study. Then the report tries to explain the working of Acunetix with proper screenshots. In the next section, few high priority vulnerabilities identified by Acunetix are described. AcuSensor and AcuMonitor technologies implemented by Acunetix is also discussed. Advantages and disadvantages of Acunetix is discussed in the next section and finally this report finishes with some recommendations and conclusion.

**Case Study: Acunetix**

Acunetix is an automated web vulnerability scanner which scans any web application or websites that use HTTP or HTTPS protocols and are accessible through a web browser. It audits the websites by identifying vulnerabilities, such as SQL injection, cross site scripting, and others.

The following table (Table 1) gives the description of the version, cost and other details of Acunetix used for this paper. Websites mainly used in the walkthrough are test websites as opposed to organisational websites. Acunetix is available in four versions, online, standard, pro and enterprise.

Table 1. *Version and platform of Acunetix used*

| Version | Acunetix 11 Trial |
|---|---|
| Platform | Windows 8 |
| Cost | Free |
| Trial period | 14 days trial |
| Test websites | http://testasp.vulnweb.com/, http://testhtml5.vulnweb.com, http://testphp.vulnweb.com |

The address of a website or web application that needs to be scanned should be added to the target. A description of the website can also be given while adding the target. Figure 1 below, is the screenshot of adding a target with description.

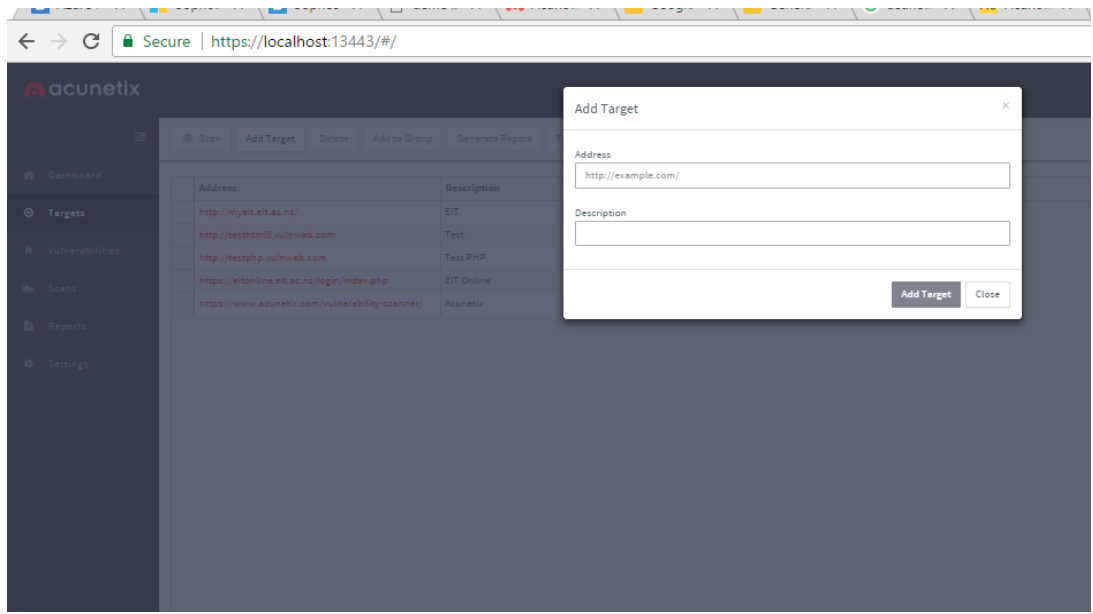

*Figure 1*. Screenshot of adding target.

After adding the target, website is ready for scanning. As seen in Figure 2, options for setting the business criticality, scan speed and others are available. Websites and web applications can also be kept for continuous scanning.

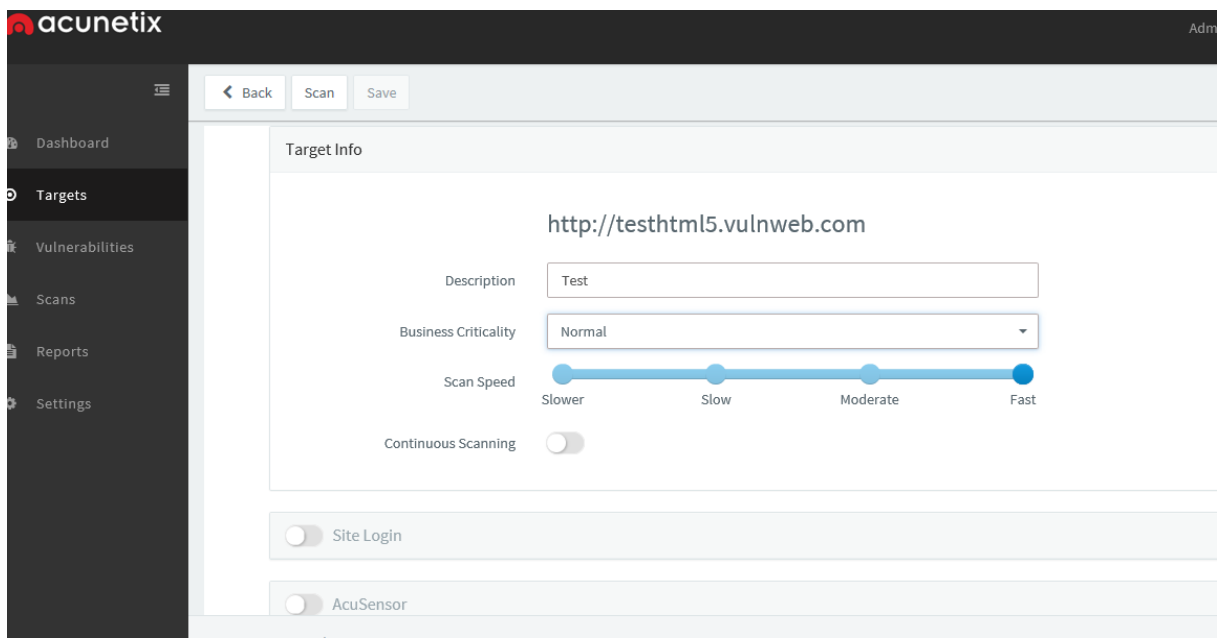

*Figure 2*. Screenshot of setting target information.

As shown in Figure 3, Acunetix offers a feature which tries to auto login to the targeted website. To accommodate this feature, two options are available. The tester can either enter the username and password manually, or a use a pre-recorded login sequence for auto login.

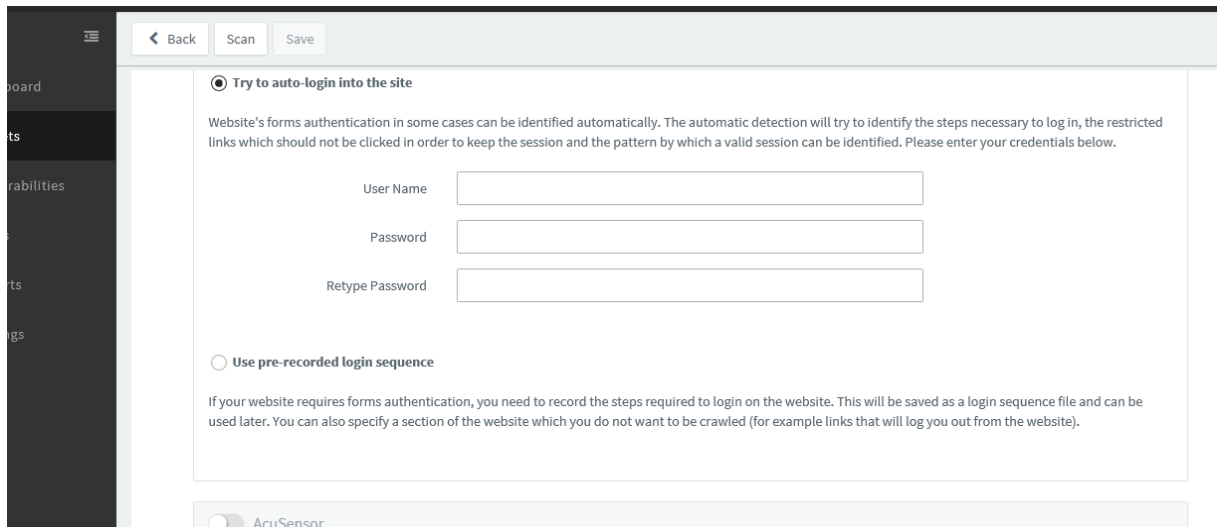

*Figure 3*. Screenshot of setting auto login option.

If the targeted web application is using PHP or .NET, the scan results can be improved by downloading and installing the proper AcuSensor. Figure 4 below shows enabling of AcuSensor.

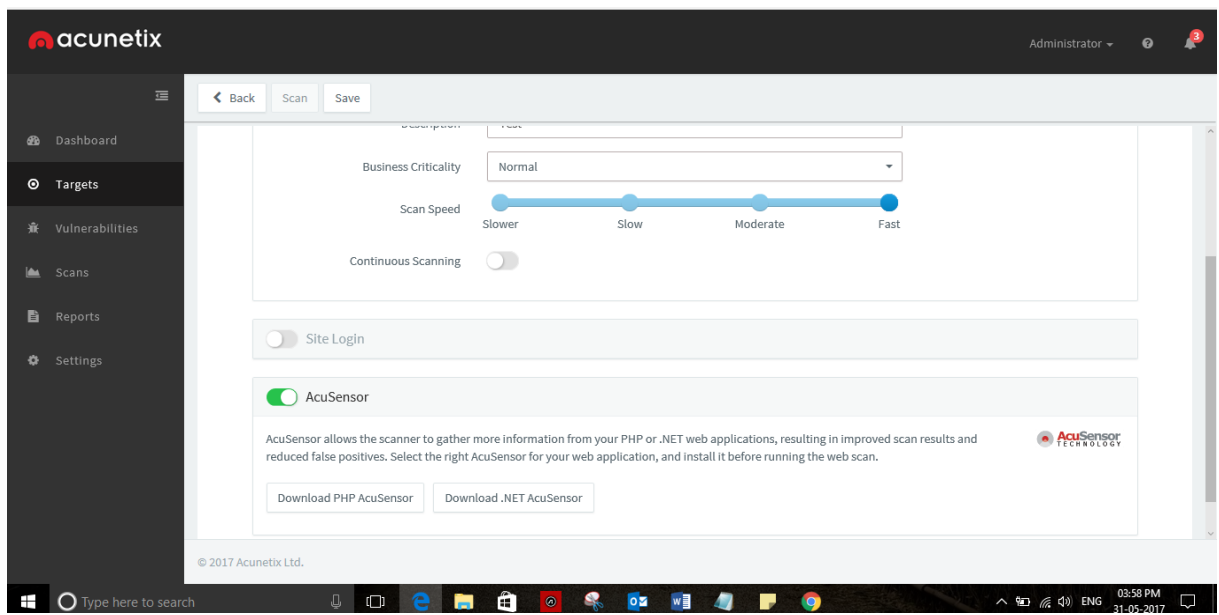

*Figure 4*. Screenshot of enabling AcuSensor.

During scanning, Acunetix provides details such as scan progress, scan duration, number of requests sent, average response time, and information about the target. It also provides the latest detected vulnerabilities and their priorities. Based on the detected vulnerabilities, Acunetix gives the overall threat level of website or web application. Figure 5 shows a screenshot of Acunetix during a scan.

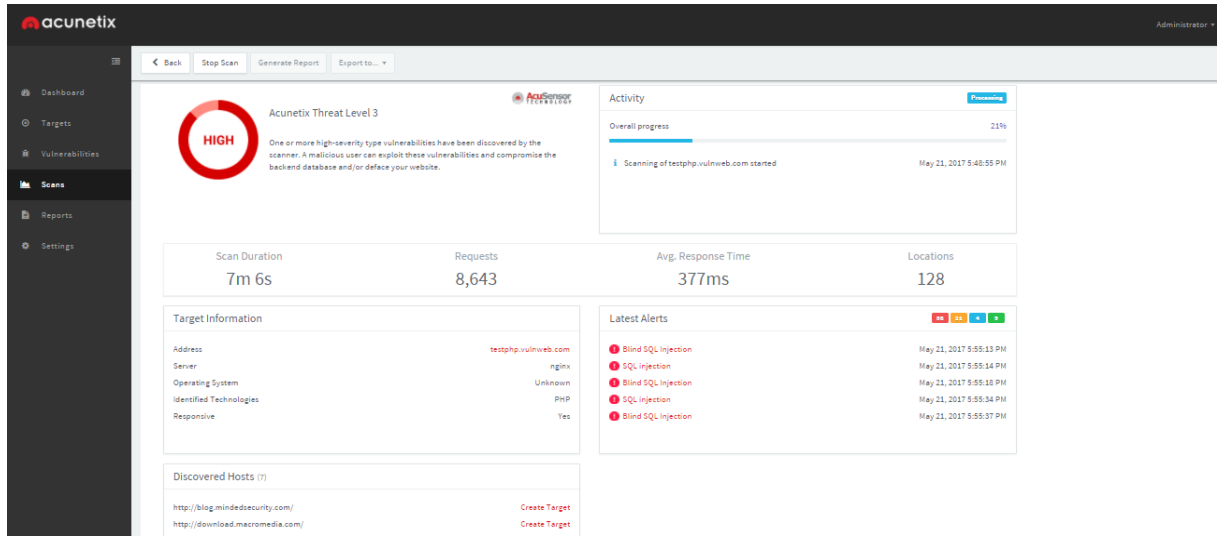

*Figure 5*. Screenshot of scanning website.

Once scanning is finished, the vulnerabilities detected by Acunetix are listed, based on the priority. It gives the name, URL, parameter, and status of the threat detected. Figure 6 shows the screenshot of scan result and Figure 7 shows the vulnerabilities identified during the scan.

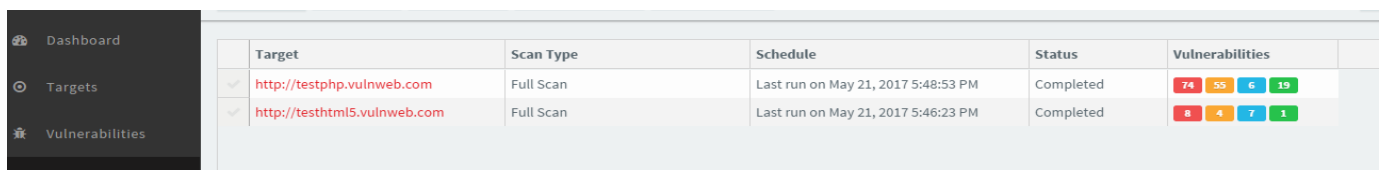

*Figure 6*. Screenshot of scan result.

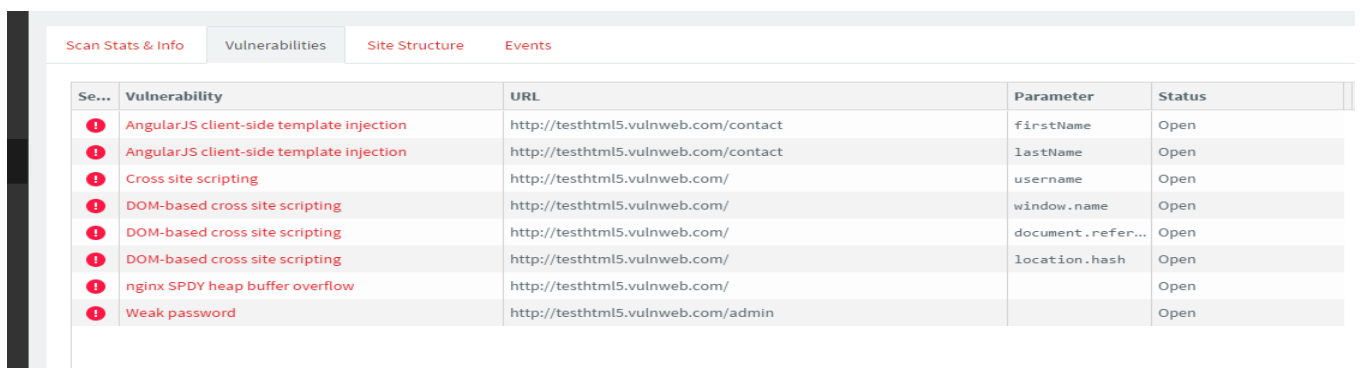

*Figure 7*. Screenshot of vulnerabilities.

Each vulnerability detected provides additional description, impact and useful tips to fix the vulnerability. It also provides the HTTP request sent, which can help in fixing and testing the particular vulnerability. Figure 8 below shows the description and impact of vulnerability "weak password".

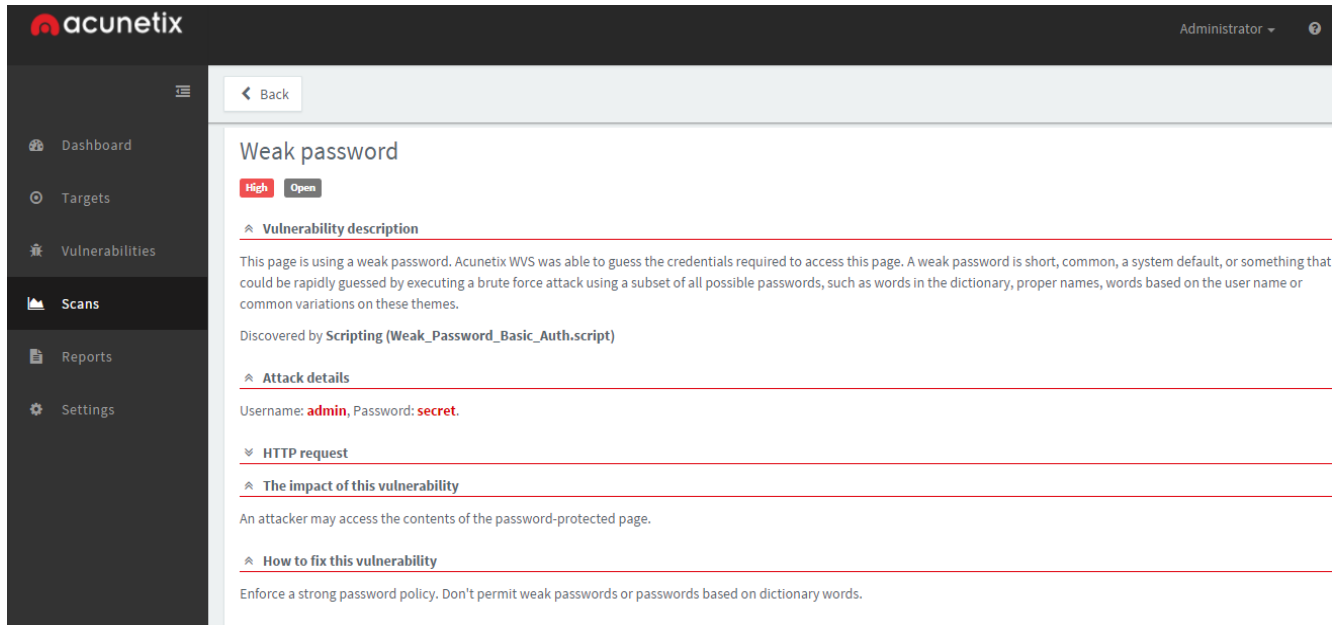

*Figure 8*. Screenshot of vulnerability description.

Scan reports can be generated based on the scan results. Reports are available in different templates, such as affected items, developer, executive summary, and also several compliance reports can be generated. All the reports generated will be available under the reports tab, and can be downloaded later. Figure 9 shows the options for generating reports.

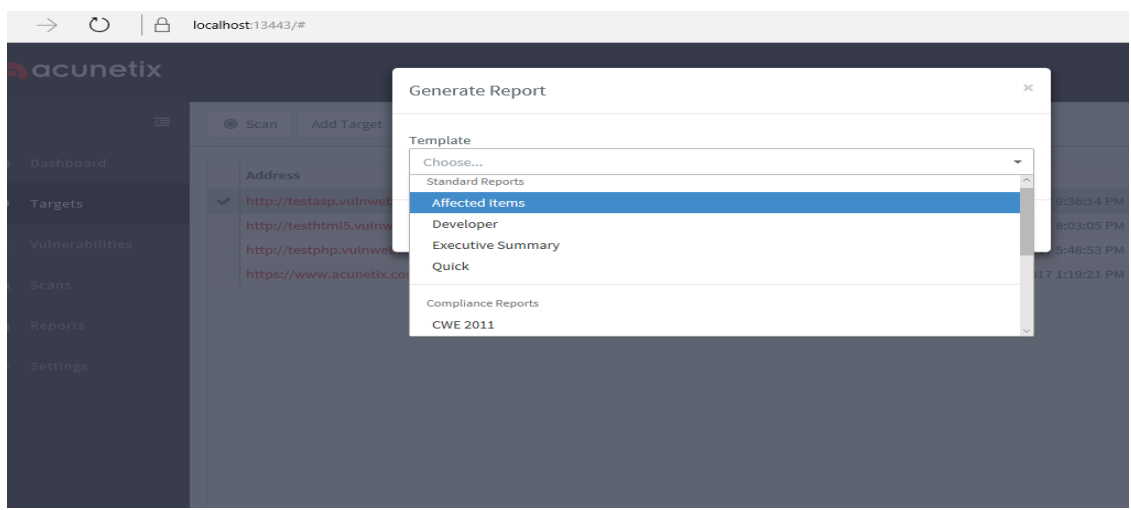

*Figure 9*. Screenshot of selecting report template.

Two scans for a single target can be compared and a comparison report is generated by Acunetix. This can be helpful to identify whether the fixes for the threats are working correctly and if those fixes do not induce more vulnerabilities. Figure 10 below shows selection of two scans on same target and "compare scans" button gets enabled.

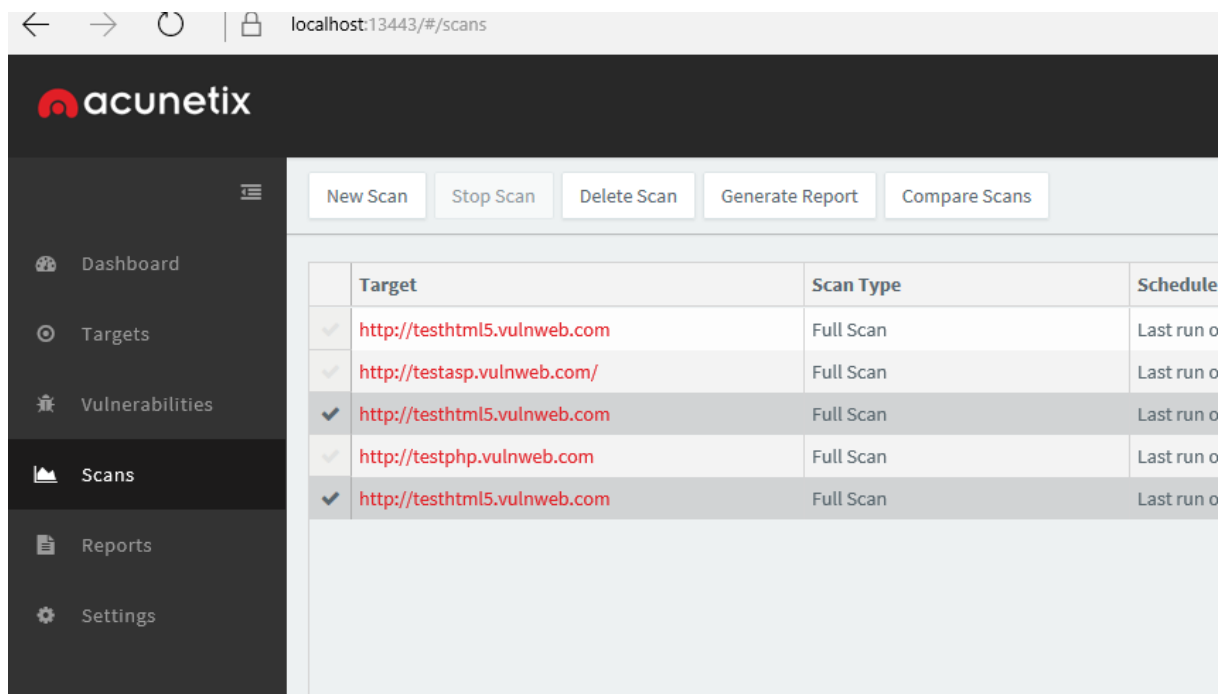

*Figure 10*. Screenshot of comparing scan results.

**Vulnerabilities**

Few high priority vulnerabilities detected by Acunetix include Cross Site Scripting (XSS), SQL injection, Blind SQL injection, and directory traversal.

XSS is inserting malicious code into a victim's web application so that, when a victim browses the web application, the malicious script code is executed (Gupta & Gupta, 2015). A hacker injects malicious codes in the dynamic websites and when the code is executed in the web browser, it changes the web pages (Jasmine, Devi, & George, 2017). The goal of an XSS attack is to get access to the client cookies or any other sensitive information, which is used to authenticate the client to the website (Jasmine, Devi, & George, 2017).

When users visit a website, their browsers send HTTP requests, in which the headers include information about their browsers and operating systems. Based on this information, the users may be directed to the mobile version of website that, along with different content, may have different vulnerabilities. This has significant implications while trying to identify XSS issues.

For this reason, Acunetix aims to crawl different versions of each website with different user agents.

SQL injection vulnerability can cause the exposure of all the sensitive data including username, password, and credit card details of a web application database (Khalid & Yousif, 2016). The SQL attacker tries to insert a part of malicious SQL commands by using special variables and inserting them in to the application. The web application in turn sends these malicious commands to the target database in the server that executes them in a different purpose using legitimate query (Abdulqader, Thiyab, Ali, & 2017). A blind SQL injection involves asking the database a series of true or false questions, in order to get closer to the vulnerable code itself. It is important to make a note here that identifying vulnerable code itself may not be sufficient for hackers on its own. Additionally, phishing is often used to obtain other user details (Erturk, 2012). These details can then be used as part of an SQL injection attack to extract unauthorized information from an online database.

Directory traversal (also known as path traversal) can be defined as an attack that "aims to access files and directories that are stored outside the web root folder" (OWASP, 2015). This vulnerability can exist in the web server or web application code. This allows the attacker to access parts of directories which are restricted, and to execute commands on the web server.

**Technologies**

Acunetix uses technologies like AcuSenor and AcuMonitor to achieve better scanning results. Acunetix AcuSensor Technology is a security technology that allows the identification of more vulnerabilities with less false positives. In addition, it indicates the exact location of the code where the vulnerability is and reports debug information. This technology combines black box scanning techniques and feedback from sensors placed inside the source code to achieve more accuracy. The screenshots below show the SQL injection (Figure 11) and PHP vulnerabilities (Figure 12) identified by the AcuSensor technology. It displays the stack trace of SQL injection threat and the file name with line number for the PHP code injection, which helps the developers to trace and fix vulnerabilities easily. It can also help developers to understand more about vulnerabilities which in turn helps them to write more secure code.

*Figure 11.* SQL Injection Reported by AcuSensor Technology

*Figure 12.* PHP Code Injection Reported by AcuSensor Technology

Another technology used by Acunetix is AcuMonitor Technology. While testing web applications, normally the scanner sends a request to a target, receives a response, analyses that response, and raises an alert based on the analysis. Some vulnerabilities do not give a response to a scanner during testing (Out-of-band vulnerability testing), and therefore, are not detectable using the "request/response" testing model. Detecting these vulnerabilities requires an intermediary service that a scanner can access. Acunetix, combined with AcuMonitor, makes automatic detection of such vulnerabilities easy. AcuMonitor detects Out-of-band SQL Injection (OOB SQLi), Blind XSS (or Delayed XSS), SMTP Header Injection, Blind Server-side XML/SOAP Injection, Out-of-band Remote Code Execution (OOB RCE), Host Header Attack, Server-side Request Forgery (SSRF), and XML External Entity Injection (XXE) automatically ("AcuMonitor," n.d.).

**Discussion**

Acunetix allows multiple scans simultaneously, but this may require more time for completing the entire scanning process. Time to scan a particular website or web application depends upon the technologies and complexity of the target website. Acunetix also allows to scan for particular vulnerabilities like XSS that can be set while starting of the scan. This is an advantage as this is much quicker compared to full scan and may help the developers to concentrate on a particular vulnerability and fix it. Reports generation is another good feature which helps to reuse the scan results. Different reports allow to get results based on specific need. Even though Acunetix reduces the false positives, scanning results may still contain false positives. Developers need to double check the scanning result and confirm that results are not false positives, which can be more time consuming. If Acunetix provides a feature to mark the false positives manually and restrict those from further scanning results, scanning results could have been better with less false positives.

Web Vulnerability Scanners (WVS) insert garbage values in the database while scanning (Suteva, Anastasov, & Mileva, 2014). WVS does automated scanning and performs several operations on a database to identify SQL injection and other threats related to the database. This can be an issue in the production environment as many garbage values are inserted with the original data.

An automated vulnerability scanner sends thousands of web requests to the web server. In order to accelerate their scanning, vulnerability scanners tend to send these requests using multiple simultaneous connections. If a web server is incapable of handling all the requests, the web server may slow down, resulting in a denial of service (Darmanin, 2014).

Automated WVS allows deep scanning. Thisi when a WVS tries to access all possible paths and links in website. This can be an issue while crawling into sensitive links. Crawling on sensitive links like delete can cause accidental deletion of some important data (Darmanin, 2014).

Some websites allow sending emails, for example, a "contact us" option in websites. While testing these websites, multiple emails can be sent to a particular address, as a mass mailing attack or mail flooding (Darmanin, 2014).

There are few solutions available for the problems discussed above. Testing in a staging environment, instead of production environment, can help prevent denial of service problems in the release environment (Darmanin, 2014). This also avoids insertion of garbage values in the production database. Denial of service in a production environment can be reduced. Using CAPTCHAs in the form for sending emails will help prevent email flooding. Furthermore, a WVS like Acunetix allows to restrict sensitive links from being crawled (Darmanin, 2014).

**Mobile Device Aspects**

Smartphones and tablets continue to pose some risks to users in terms of web security. These devices typically handle different wireless connections, cached or saved passwords, and notes and emails containing private information. As these bits of data are stored on mobile devices, they may be exposed to being stolen through web browsers. The education sector comprises the largest and growing groups of users of mobile devices. School teachers, employees and students constantly rely on their websites and online learning tools, and these are frequently facing security threats (Levin, 2017). More than ever, the users of educational information systems need to be more alert and better trained in security matters. IT support services for schools need to build their protective capacity and carry out new security practices. Mobile and web security aspects also need to be covered by analysts and developers of educational applications by making security a crucial requirement and design principle (Erturk, 2013).

**Conclusions**

Web Vulnerability Scanners (WVS) help to speed up the website and web application vulnerability testing process. Acunetix is a popular automated vulnerability scanner, which not only identifies vulnerabilities, but also gives suggestions for solving these vulnerabilities. AcuSensor technology also focuses on reducing the reporting of false positives for websites based on PHP and .NET technologies. utilizing new behavioural analysis techniques in the future can make Acunetix even more accurate. Using of Acunetix or a similar WVS is very useful for many web developers since this type of tool makes the detection of threats easier for security novices. Using a trial version can be helpful in make the decision regarding how long a period to choose for the paid subscription in the future. Different plans help the users to choose a suitable subscription option according to their needs. Website and application developers can use WVS during their development and testing to ensure a very secure web application before it is released into the production environment. This case study is helpful for orienting students with the basics of a WVS, with Acunetix as an example. Whereas some security solutions are geared towards examining organisational websites, other solutions may focus on managing the organisation's mobile devices. It is useful for both app designers and for IT support to scan mobile devices to identify the security vulnerabilities and to highlight insecure apps (Revankar, 2015). Further technical studies can be done to compare different vulnerability scanners, their effectiveness, and their peculiar strengths, which in turn would help developers choose an appropriate WVS for each web application.